# Effect of Scaffolding on Helping Introductory Physics Students Solve Quantitative Problems Involving Strong Alternative Conceptions

Shih-Yin Lin

*Department of Physics, National Changhua University of Education, Changhua, 500, Taiwan*

Chandralekha Singh

*Department of Physics and Astronomy, University of Pittsburgh, Pittsburgh, PA 15260, USA*

It is well-known that introductory physics students often have alternative conceptions that are inconsistent with established physical principles and concepts. Invoking alternative conceptions in quantitative problem-solving process can derail the entire process. In order to help students solve quantitative problems involving strong alternative conceptions correctly, appropriate scaffolding support can be helpful. The goal of this study is to examine how different scaffolding supports involving analogical problem solving influence introductory physics students' performance on a target quantitative problem in a situation where many students' solution process is derailed due to alternative conceptions. Three different scaffolding supports were designed and implemented in calculus-based and algebra-based introductory physics courses involving 410 students to evaluate the level of scaffolding needed to help students learn from an analogical problem that is similar in the underlying principles involved but for which the problem solving process is not derailed by alternative conceptions. We found that for the quantitative problem involving strong alternative conceptions, simply guiding students to work through the solution of the analogical problem first was not enough to help most students discern the similarity between the two problems. However, if additional scaffolding supports that directly helped students examine and repair their knowledge elements involving alternative conceptions were provided, e.g., by guiding students to contemplate related issues and asking them to solve the targeted problem on their own first before learning from the analogical problem provided, students were more likely to discern the underlying similarities between the problems and avoid getting derailed by alternative conceptions when solving the targeted problem. We also found that some scaffolding supports were more effective in the calculus-based course than in the algebra-based course. This finding emphasizes the fact that appropriate scaffolding support which is commensurate with students' prior knowledge and skills must be determined via research in order to be effective.

# I.    INTRODUCTION

Helping students become adept at both quantitative and qualitative problem solving is an important goal of many physics courses [1,2]. In order to solve a problem appropriately, students must identify the physics concept(s) and principle(s) relevant to the problem, and apply their knowledge in an appropriate manner to reach the targeted goal. However, research suggests that many students have alternative conceptions in physics that are different from the scientifically established ones [3-9]. These conceptions may come from students' interpretation of their past experiences [10], and they can influence students' learning of the scientifically established concepts. Prior research has shown that some alternative conceptions are so robust that many students still hold the same conceptions after traditional instruction which may not give students an opportunity to examine, repair and build a good knowledge structure [11]. As a result, students may have a fragmented knowledge structure in which both the alternative conceptions and the scientifically established conceptions that students learn in their physics classes coexist [8,12]. However, students may not necessarily notice the conflicts between them. While solving problems, students may use different problem solving approaches to solve problems that are considered very similar by experts based upon deep features, depending on the knowledge elements that were perceived to be relevant in a given context by the student [13-17]. If alternative conceptions are invoked while solving a problem, students are likely to use a problem solving approach involving physics principles and concepts not appropriate in the given situation.

In order to help students solve problems involving alternative conceptions, additional instructional scaffolding supports may be useful. While the research literature on alternative conceptions has researchers expressing somewhat different views [15,18-30] including how the details of conceptual change processes occur [8,12,15,21,27,29-36], most instructional strategies developed to facilitate conceptual change involve helping students recognize existing conceptions (e.g., by making students aware of the inconsistencies or conflict associated with their conceptions and physics concepts) and guiding them through a series of activities to reflect on, refine, and reorganize their knowledge elements [20,37-42]. This study builds on instructional strategies suggested by these lines of research and investigates how different scaffolding supports involving analogical problem solving impact introductory physics students' quantitative problem solving performance in a situation involving strong alternative conceptions. The level of scaffolding support needed to assist students in solving the quantitative problem will be explored.

We have chosen analogical problem solving [43-48] as the framework for instructional scaffolding used in this study for several reasons. First, analogy has long been a strategy adopted by many instructors in the classrooms. With the help of analogies, students can blend in knowledge from situations that they are already familiar with to construct understanding in a new, unfamiliar situation [11,49]. It is also common for students to solve a new problem by drawing analogy with problems they already know how to solve and applying similar reasoning strategies from one problem to another. Moreover, we hypothesize that through careful design, the analogy between two problems can be used to help students discern the inconsistency between their alternative conceptions and the scientifically established conceptions, which can be useful for fostering a conceptual change [8,27,31]. For example, if students are presented with two problems, both of which can be solved using the same principles of physics but one problem (problem A) involves a situation in which students are likely to use their alternative conceptions to solve the problem while the other (problem B) does not, students can be guided to explore two possible ways to solve problem A: one derived from the similarity between the two problems, and the other based on their alternative notion. The conflict found between these two possible solutions can be used to guide students to evaluate their alternative conceptions while solving the quantitative problem and to contemplate whether those conceptions are applicable in the given context. We hypothesize that if students realize that the conceptions they initially thought were valid in a context are not appropriate in that situation, the similarities between the analogical problems can be used to help them reorganize and repair their knowledge structure.

We note that if introductory students are guided to explore their alternative conceptions in depth, they may also be able to discern incoherence in their reasoning of problem A without the use of an analogical problem involving the same underlying physics principle that does not involve such conceptions (problem B). However, it is unlikely that students would use a different conception (and not use their alternative conceptions) to solve problem A unless they are provided appropriate guidance and support to explore a new way of solving it that focuses on helping them re-examine and repair their relevant knowledge structure. Providing students with the analogical problem B at this point can be one strategy to help them discern a different way to approach problem A, and increase the likelihood for students solving problem A successfully. Such strategies are also aligned with Brown and Clements' suggestion of how "bridging analogies" [11] may be used to foster the application of a concept in the targeted situation which may involve an alternative conception. In particular, Brown and Clement suggest that by starting with an anchoring example (for which students' conception is aligned with the scientifically established concepts), analogies can be used to gradually "bridge" students' conception from the anchoring example to the targeted situation. Problem B in the strategies described above could play a similar role to the anchoring example in helping students re-craft their approaches to solving problem A and re-examine and repair their knowledge structure. As prior research efforts have found mixed success in the use of analogy [11,50-54], it is important to study the level of scaffolding needed to help students successfully solve the targeted problem, especially when strong alternative conceptions are involved. This paper describes an investigation along these lines in a quantitative problem solving task and examines the effect of different scaffolding supports provided to students.

In the following subsection of the introduction, we introduce the quantitative problem used in this investigation and students' common alternative conception related to this problem. Results from previous studies that shed light on the current investigation is also summarized.

### A. Students' alternative conception related to the quantitative problem posed

Figure 1 shows the quantitative problem used in this study, which is about a car at rest on an inclined plane. Students are asked to solve for the frictional force acting on the car. Prior research suggests that many students have the conception that the magnitude of the static frictional force ($f_s$) is always equal to its maximum value, the coefficient of static friction ($\mu_s$) times the magnitude of normal force ($F_N$) [55]. Students often rely on this conception when they are asked to solve for the frictional force acting on an object held at rest. However, this conception is not valid in problems such as the one shown in Figure 1 because the maximum value of static friction exceeds the actual frictional force needed to hold the object at rest. The process of correctly solving the problem shown in Figure 1 involves using Newton's 2$^{nd}$ Law and realizing that the static frictional force is equal to mg cos 30⁰, the component of gravitational force parallel to the surface of the inclined plane. Substituting the values given in the problem, the correct magnitude of the frictional force is 7500 N. However, if students calculate the static frictional force using $f_s = \mu_s F_N$, they will arrive at $f_s = 11691$ N, which is greater than the actual magnitude of the frictional force. A previous study [55] shows that when the same friction problem was given to a group of introductory physics students in a multiple-choice format, only 20% of the students selected the correct answer. About 40% of the students selected an incorrect response corresponding to $f_s = \mu_s F_N$. On the other hand, when a similar problem (shown in Figure 2) involving tension instead of friction was given to students, 72% of the students were able to solve for the tension force correctly by using Newton's Second law [55]. This study suggests that the difficulty many students have with the friction problem does not result from a lack of procedural facility with Newton's 2$^{nd}$ law, which, for both the tension and friction problems suggests that the net force acting on an object should be zero when it is at rest, or from an inability to decompose forces correctly. Rather, the difficulty often comes from students' alternative conception about friction, which hinders their ability to solve the friction problem (Figure 1) by applying Newton's 2$^{nd}$ Law in the direction parallel to the inclined plane.

A car which weighs 15,000 N is at rest on a 30⁰ incline as shown. The coefficient of static friction between the car's tires and the road is 0.90, and the coefficient of kinetic friction is 0.80. Find the magnitude of frictional force on the car. Note: These trigonometric results might be useful: sin 30⁰=0.5, cos 30⁰=0.866 to three places.

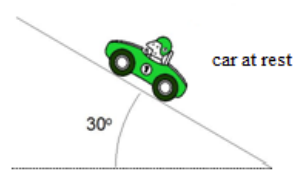

**Figure 1. Friction problem used in the study**

A car which weighs 15,000 N is at rest on a frictionless 30⁰ incline as shown. The car is held in place by a light strong cable parallel to the incline. Find the magnitude of tension force in the cable. Note: These trigonometric results might be useful: sin 30⁰=0.5, cos 30⁰=0.866 to three places.

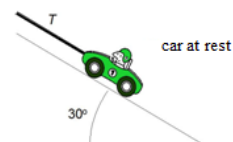

**Figure 2. Tension problem (the analogical problem)**

**B.   Results from previous studies that shed light on the current study**

Since prior research suggests that many students are able to solve the tension problem (Figure 2) correctly, it is possible that using the tension problem as a scaffolding support can help students solve the friction problem correctly if they are guided to perform analogical problem solving between these two problems. One question that has been explored in prior studies is whether the tension problem alone is enough to improve students' performance on the friction problem, or whether other scaffolding supports are needed for students to examine and repair their knowledge structure and help them solve the friction problem (in which many students do not apply Newton's 2$^{nd}$ Law along the direction parallel to the inclined plane to solve the problem). As suggested by prior studies [55,56], simply providing two problems as a pair (e.g., by posing them back to back) does not help students much because such intervention are too implicit and students do not necessarily discern the connection between the two problems. Moreover, students' alternative conception of static friction may deter analogical reasoning and prevent transfer of knowledge between these two problems. In particular, the problem solving process of some students suggested that even if students obtained the correct numerical value for static frictional force, they still held the alternative conception about friction [56]. For example, a student attempted to solve for static frictional force by (1) first setting up an equation: $\mu_s F_N - mg \cos 30^o = 0$, (2) plugging in values for $\mu_s$ and $mg \cos 30^o$ to the equation above to solve for the value of $F_N$, and (3) using the values of $\mu_s$ $and$ $F_N$ to solve for the frictional force. We note that although this student obtained the correct numerical value for static frictional force, he had obtained an incorrect numerical value for the normal force. Overall, the prior studies suggest that simply pairing problems and expecting students to perform analogical problem solving between the problem pair was not very effective in helping students repair their knowledge structure and solve problems involving alternative conceptions correctly. More scaffolding supports are required to help students process through the analogy between the two problems deeply and to examine the applicability of their alternative conception in the given situation.

In light of these prior studies [55,56], three different interventions were designed in this study to provide students with additional support and to investigate the level(s) of scaffolding needed to help students solve the friction problem appropriately. We discuss the designs of these interventions in the Methodology section, and report the findings in the Results and Discussion section.

# II.   METHODOLOGY

## A.   In-class study

To evaluate the effect of different scaffolding supports in helping students solve a problem involving a strong alternative conception, the friction problem (shown in Figure 1) with different scaffolding supports was implemented in two introductory physics courses in a recitation quiz setting. One hundred and eighty three students from a calculus-based introductory mechanics course and 227 students from an algebra-based introductory mechanics course were involved in this study. The study was implemented in each course after the concepts of force equilibrium and frictional forces had been discussed in the lecture. Although the instructor for the calculus-based course was different from the instructor for the algebra-based course, both instructors emphasized the inequality of static friction in their classes. For example, the instructor for the algebra-based course demonstrated the inequality with an inclined plane demonstration in which the angle of inclination could vary until the block on the incline started sliding, and the instructor for the calculus-based course discussed the inequality using an example of a block on a horizontal surface, and explained how the magnitude of the frictional force and the motion of the block are impacted if the applied force on the block is linearly increased from zero.

The exact time of the implementation was determined by each instructor based on how they believed the quiz in this study best fit their course schedule. In the calculus-based course, the study was implemented as a quiz after the topic of frictional force was discussed in lectures and students submitted their homework on this topic but before an exam on relevant topic took place. In the algebra-based course, since part of the lecture about frictional force was completed in the week right before a midterm exam (that midterm exam included frictional force), the problem in this study was given as a quiz that took place in the week right after the midterm exam. Since problems about frictional force were included in the midterm exam, the concepts related to friction would still be fresh in students' minds when they took the recitation quiz one week later.

To evaluate the effects of different interventions, in each course, students were divided into four groups – one comparison group and three intervention groups - based on different recitation classes. Investigation of student performances on the Force Concept Inventory (FCI) [57] administered at the beginning of the semester and their scores on the final exam suggested that in each course, students in all four groups were comparable on these measures. There was no statistically significant difference between different groups in terms of the FCI scores or the scores on the final exam.

Students in the comparison group were asked to solve the friction problem in a quiz on their own. No tension problem or any other scaffolding support was provided. Examining the performance of this group of students can help us understand what students in this population can accomplish when no scaffolding support is provided. Students in the three intervention groups received the tension problem (shown in Figure 2) and its solution to help them solve the friction problem. They were instructed to learn from the solution of the tension problem provided to them, draw an analogy between the tension and friction problems by explicitly pointing out the similarities between them, and write down how they can take advantage of the tension problem provided to solve the friction problem. In addition to these instructions, different additional scaffolding supports were implemented in different intervention groups to help students process through the analogy deeply and/or to contemplate the applicability of associating the static frictional force with its maximum value. A summary of the different scaffolding supports implemented in each intervention group is presented in Table 1.

Before describing the details of each intervention, we note that different types and levels of scaffolding supports were chosen so that the effects of different potentially helpful interventions can be investigated. These interventions were designed based upon a cognitive task analysis [58,59] from an expert perspective in which experts reflect on their own underlying thought processes when reasoning about friction and the difficulties that were identified in the previous studies about why novices struggle to solve the problem. There is more than one difference between the scaffolding supports provided in different interventions, because it is not clear a priori what scaffolding supports would be effective, and we wanted to explore a set of scaffolding supports that potentially could be helpful. Although there are other promising scaffolding supports, we had to select only three for the three recitation classes. Further hypothesis for why the researchers decided that the three interventions selected could potentially be beneficial is given later when describing each intervention. We also note that since the prior studies suggest that students may use Newton's 2nd Law in the direction parallel to the inclined plane to solve for friction while simultaneously claiming that $f_s = \mu_s F_N$ is valid [56], in order to better explore students' reasoning difficulties about static frictional force, students in our current study were explicitly asked to solve for both (1) the static friction and (2) the normal force in the friction problem.

**Table 1. Summary of different interventions used in the study. The time[1] provided to students to complete each task is included in the parentheses. We note that students in all groups had enough time to complete the task.**

| group | Students were asked to…. |
|---|---|
| Comparison group (No intervention) | Solve the friction problem (problem involving an alternative conception) on their own in 15 minutes. No scaffolding support is provided. |
| Intervention group 1 | (a) First learn from the tension problem (i.e., the analogical problem that does not involve the alternative conception) whose solution was provided (10 minutes)<br>(b) Return the solution to the tension problem<br>(c) Solve both the tension problem and friction problem (20 minutes) |
| Intervention group 2 | (a) First solve the friction problem on their own. Students were also asked to predict whether the static frictional force should be larger or smaller if the same car is at rest on a steeper incline, and to compare their calculation with their prediction. (10 minutes)<br>(b) Learn from the solution to the tension problem and redo the friction problem a second time (with the solved tension problem in their possession) (20 minutes) |
| Intervention group 3 | Learn from the solution to the tension problem and then solve the friction problem (with the solved tension problem in their possession). Students were also asked to explain the meaning of the inequality in $f_s \leq \mu_s F_N$ (25 minutes) |

1.    *Details of the different scaffolding supports used in three different intervention groups*

In order for the analogical problem solving activity to successfully help student solve the friction problem, students must understand that solving the tension problem involves the same

---

[1] The time provided in each group was determined using the following guidelines: (a) Our prior experiences suggested that students were generally able to complete the tension problem or the friction problem within a 10-minute quiz, and many of them did not need the full 10-minute period to work on it. Based on these prior experiences, at least 10 minutes were given to students for each of the following: solve the tension problem, solve the friction problem, or learn from the solution to the tension problem (b) Due to the length of the recitation classes, the quiz should be kept within 30 minutes long. If the time for the entire quiz did not exceed this limit, another 5 minutes were added to the corresponding intervention group so that students could have more time to work on the task if they want to

underlying physics principle (i.e., Newton's 2nd Law) as the friction problem, and that these two problems can be solved using a similar approach. All of the different scaffolding supports used in the three interventions were designed with the intention to facilitate this process.

In particular, students in the intervention group 1 were asked to take ten minutes to learn from the solution to the tension problem provided to them before they received the friction problem. They were explicitly told at the beginning of the task that after ten minutes, they had to return the solution to the instructor, and then they would be given two problems to solve: one of them would be the exact same problem they just browsed over (i.e., the tension problem), and the other one would be similar (i.e., the friction problem). We hypothesized that by asking students to display how to solve the tension problem again on their own, they will process through the concepts they learned from the tension problem in more depth than those in the prior study [56]. Moreover, solving the tension problem on their own again may prime students to apply a similar approach later when solving the friction problem. For example, if students adopt the problem solving approach used in the solved problem and set up both problems by drawing the free-body diagrams first, they may be more likely to realize that these two problems have the same free-body diagram, and that the friction in one problem should have the same magnitude as the tension in the other problem. We note that intervention 1 focused on the use of similarities between the tension problem and target problem to foster the appropriate problem solving approach without explicitly providing scaffolding to help students examine and repair their fragmented knowledge involving alternative conception. Since this intervention did not directly focus students' attention on the alternative conception they commonly have in this context, students were expected to recognize the conflict between their alternative conception and the implication of the problem similarities on their own. Examination of student performance in this group can help us understand what students are able to achieve with this level of scaffolding provided.

Unlike intervention 1, interventions 2 and 3 were designed to address students' alternative conceptions about the static frictional force in a more direct manner by providing them with additional scaffolding tasks that may help them recognize the incoherence in their approach involving alternative conception. In particular, students in intervention 2 were asked to make a qualitative prediction about the magnitude of the static frictional force (whether it's larger or smaller) when the same car is at rest on a steeper inclined plane (with the same static coefficient of friction) based on their daily experience. They were also explicitly instructed to quantitatively calculate the magnitudes of the frictional force acting on the car with two different degrees of inclination and compare their quantitative result with their qualitative prediction to check for consistency. Our hypothesis based upon a cognitive task analysis from an expert perspective was that students could reason from their daily experience that it's more difficult to stand still on a steeper inclined plane; therefore, a larger frictional force would be required in order for the same car to stay at rest on a steeper incline. However, if a student used $f_s = \mu_s F_N = \mu_s mg \cos \theta$ (where $\theta$ is the angle of inclination) to calculate the magnitude of the frictional force, there would be a conflict because as the degree of inclination increases, the normal force decreases, making the frictional force calculated in this manner smaller. We hypothesized that if students notice this conflict but do not know how to resolve it on their own, the similarities between the tension problem and the friction problem (together with the solution to the tension problem provided) can help students recognize a new approach to solve the friction problem (i.e., using Newton's 2nd law along the direction parallel to the plane). We also hypothesized that if students are asked to solve the friction problem and contemplate those additional questions in this intervention first before the tension problem with its solution is provided, students would be more likely to notice the deficiency in their conceptions about static friction. After recognizing the conflict, they may pay more attention to the similarities between the problems and benefit more from the tension problem provided to repair their knowledge structure and solve the friction problem correctly. Therefore, students who received intervention 2 were asked to take the first 10 minutes to solve the friction problem set (which, in addition to the original friction problem, included sub-problems asking for a qualitative prediction and a quantitative calculation of the magnitude of $f_s$ on a steeper incline as well as a consistency check) on their own before the solved tension problem was provided as a scaffolding tool. After students completed the problem set the first time, they turned in their first solution to the instructor, and then they were given the tension problem with its solution. With the solved tension problem in their possession, they were asked to learn from the tension problem and solve the friction problem set (again the extra sub-problems are included) a second time. The design of scaffolding support in this intervention was inspired by cognitive theory [8,12,60,61] which suggests that cognitive conflict can be useful for helping students learn concepts, examine and repair their knowledge structure and build a better understanding. The complete instructional materials used in intervention group 2 can be found in the auxiliary materials.

A different scaffolding support aimed at guiding students to examine the applicability of the relation $f_s = \mu_s F_N$ was implemented in intervention 3. Students who received intervention 3 were provided with the solved tension problem along with the friction problem. In addition to the instruction asking them to discuss the similarity between the two problems before solving for the frictional force, they were also asked to explain the meaning of the inequality in $f_s \leq \mu_s F_N$ and discuss whether they can find the frictional force on the car in the friction problem without knowing $\mu_s$. Based upon a cognitive task analysis, we intended that this additional scaffolding by questioning may provide a direct hint to students to resolve the "conflict" if they are able to recognize the similar roles played by the tension and the friction in these problems but are concerned about the fact that the equation $f_s = \mu_s F_N$ does not yield an answer for friction which has the same magnitude as the tension. In order to increase the possibility of students discerning this discrepancy between the free-body diagram and $f_s = \mu_s F_N$ they used for static friction, after solving for the frictional force, the last part of the quiz explicitly asked students to solve for the magnitude of the normal force "using the component of force perpendicular to the inclined plane" (and check that the calculated normal force is consistent with what they obtained previously if the normal force was used to solve for friction). We hypothesized that if the students used convoluted reasoning as discussed earlier [56] and set $mg\,sin\,\theta - f_s = mg\,sin\,\theta - \mu_s F_N = 0$ when solving for friction, this additional hint and instruction provided may scaffold the problem solving process and help them repair their knowledge structure.

## *2.    Rubric*

To evaluate how the scaffolding supports affect student performance on the friction problem, the problem was graded using a rubric developed iteratively by deliberation between the two researchers. When two researchers scored independently a sample of at least 10% of the students using the final version of the rubric, an inter-rater reliability of more than 90% was achieved. For intervention group 2, in which students were asked to calculate the magnitude of friction with different angles of inclination, in the few cases for which there was a discrepancy between the two calculations performed by a student, the score was assigned based on the 30 degree case, which is the same as the grading of the students in the other intervention groups.

Table 2 summarizes the rubric used to score students' performance in calculating the friction force, which had a full score of 10 points. The rubric was constructed based on students' different problem solving approaches and their common mistakes. Different approaches were assigned different maximum scores because they represented different levels of understanding. For example, the maximum score a student could receive if she/he correctly used Newton's 2nd Law in the equilibrium situation ($\sum F = 0$) in the direction parallel to the inclined plane was 10 points, while a student who used $f_s = \mu_s F_N$ to solve for friction could earn a maximum score of only 5 points. If the students used $\sum F = 0$ in the direction parallel to the inclined plane and came up with the correct value ($f_s = mg \sin \theta$) for the calculation of friction, but their answers to other part(s) of the problem indicated that they still related the static friction with its maximum value (for example, first finding $f_s = mg \sin \theta$ correctly but then incorrectly using $f_s = \mu_s F_N$ to solve for the normal force in the next sub-problem), they were classified as having the same alternative notion as students using the $f_s = \mu_s F_N$ approach and the maximum score they could receive was 5 points. In other words, when grading student performance, the researchers first determined which category a student's solution fell into. If the equation $f_s = \mu_s F_N$ was employed in any part of the solution and there was no clear indication of the inapplicability of this equation in the given situation, then students were classified under the "$f_s = \mu_s F_N$" approach category with a maximum score of 5 points. These students would get a score of "5 minus whatever points they should be taken off due to other mistakes (such as those common mistakes listed in Table 2)". If a student's solution fell into the "$\sum F = 0$ in the direction parallel to the inclined plane" category, she would get a score of "10 minus whatever points were taken off due to other mistakes listed in Table 2". The researchers never used both part 1 and part 2 of the rubric to grade the same student.

Under each approach, the common mistakes students made and the corresponding points taken off are listed in the rubric. For example, students lost point(s) for decomposing the force incorrectly or for confusing weight with mass. We note that the common mistakes listed in Table 2 that students typically made were similar across different approach categories. Each student was penalized only once for each type of mistake made. In addition to the final scores students received based on the rubric, the percentages of students who adopted a particular problem solving approach in different intervention groups were also recorded by researchers for analysis.

**Table 2. Summary of the rubric used to score students' performance in calculating the frictional force.**

| Problem solving approach | Maximum score | Common mistakes (Points taken off) |
|---|---|---|
| Using $\sum F = 0$ (i.e. $f_s - mg\sin\theta = 0$) | 10 | * Minor mistake in the force decomposition such as confusing sine with cosine and ending up with $f_s = mg\cos\theta$ (-1)<br>* More serious mistake in the force decomposition such as treating $mg$ as the component of a force $f_{parallel,down}$ that should cancel $f_s$ (instead of treating $f_{parallel,down}$ as the component of $mg$) and ending up with $f_s = mg/\sin\theta$ (-2)<br>* Confused weight with mass and multiplied the weight by an additional g=9.8m/s$^2$ (-1) |
| $f_s = \mu_s F_N$ | 5 | * Minor mistake in the force decomposition such as confusing sine with cosine and ending up with $F_N = mg\sin\theta$ (-1)<br>* More serious mistake in the force decomposition such as thinking $F_N = $ mg (-2)<br>* Confused weight with mass and multiplied the weight by an additional g=9.8m/s$^2$ (-1) |
| $f_k = \mu_k F_N$ | 3 or 4 | |
| Combined $\mu_s$ and $\mu_k$ (e.g. $f_{friction} = \mu_s F_N + \mu_k F_N$) | 2 | |
| $f_s = \mu_s F_N - mg\sin\theta$ | 2 | |

## B.  Out-of-class study: Interviews

In addition to 410 students from two introductory physics courses who participated in the study in a recitation quiz, ten additional introductory physics students from other equivalent introductory physics classes were recruited for semi-structured interviews. They were invited to participate in a one-on-one interview session in which they were asked to solve the problem with one of the interventions used in the quiz study while thinking aloud. They were not disturbed during this task except being asked to talk aloud if they became quiet for a while. After they completed the task to the best of their ability, the interviewer asked them for clarification of points they had not made clear earlier to understand the cognitive mechanism involved in problem solving. Moreover, if they had not been able to solve the friction problem correctly with the scaffolding provided, the researcher would then provide additional support to the students in order to help them solve the problem correctly. The goal of the interview part of the study was to help us better understand students' thought processes while solving the problem including their reasoning related to $f_s = \mu_s F_N$, and to examine additional scaffolding students may need (if any) to help them solve the problem successfully. The details of the interviews will be discussed in Section III.B.

Before proceeding to the results section, we suggest that the readers look through the scaffolding materials provided in the auxiliary materials and make a prediction about the effects of scaffolding support in each intervention.

## III.    RESULTS AND DISCUSSION

### A.    Data obtained from recitation quizzes in introductory physics courses

*1.    Students in the intervention groups generally performed better than those in the comparison group. Also, some interventions were more helpful than others.*

Table 3 presents students' average scores on the friction problem in the calculus-based and algebra-based courses. The p-values for the comparison of students' performance between different groups are listed in Table 4. Table 3 and Table 4 indicate that in both courses, students who received the scaffolding supports significantly outperformed (with p-values less than 0.05) the comparison group students who did not received scaffolding support, except for the calculus-based intervention group 1 students whose performance was not better enough to be significantly different from the corresponding comparison group (p=0.06). Comparing the effects of different interventions in our study, we found that in the calculus-based course, the scaffolding supports of interventions 2 and 3 appeared to help students perform slightly better than intervention 1 (although not necessarily statistically different). Students in the intervention groups 2 and 3 on average achieved a score of 7.2 or 7.1 out of 10, while students in the intervention group 1 had an average score of 5.9. In the algebra-based course, the p-values show that intervention group 2 students, who achieved an average score of 6.8, performed significantly better than students in intervention groups 1 and 3, whose average scores were 5.1 and 5.0, respectively. There was no significant difference between the latter two groups. Combining the data from these two courses, we found that intervention 2 appears to always be one of the most effective interventions for both algebra- and calculus-based students.

**Table 3. Students' average scores for magnitude of friction (out of 10 points) in the calculus-based and algebra-based courses. The numbers of students in the comparison (Comp) group and each of the three intervention (Intv) groups are shown in parentheses. In the intervention group 2, students' performance both before and after they received the scaffolding was examined and the average normalized gain[2] was calculated.**

| | Comp | Intv 1 | Intv 2 | | | Intv 3 |
| --- | --- | --- | --- | --- | --- | --- |
| | | | Before | After | Normalized gain | |
| Calculus | 4.4 (38) | 5.9 (34) | 4.3 | 7.2 (72) | 0.51 | 7.1 (39) |
| Algebra | 3.1 (47) | 5.1 (63) | 4.0 | 6.8 (51) | 0.47 | 5.0 (66) |

**Table 4. The p values (from ANOVA) for the comparison of students' performance on the friction problem between different groups in the calculus-based and algebra-based courses.[3] The algebra-based course is indicated by the shaded background.**

| | Comparison | Intervention 1 | Intervention 2 | Intervention 3 |
| --- | --- | --- | --- | --- |
| Comparison | -- | 0.060 | 0.000 | 0.000 |
| Intervention 1 | 0.005 | -- | 0.057 | 0.112 |
| Intervention 2 | 0.000 | 0.009 | -- | 0.904 |
| Intervention 3 | 0.006 | 0.878 | 0.006 | -- |

[2] The normalized gain is defined by the change in the score divided by the maximum possible score for improvement.

[3] Since the scores students received were influenced by the point deduction for the common mistakes shown in the rubric, the comparison group may potentially have a disadvantage because this group did not receive the solution to the tension problem, which included some hints that may help prevent these mistakes. However, we have also compared the scores between different intervention groups when students were NOT penalized for making these common mistakes. The results indicate that while the averages score of the comparison group increased slightly more than those of the intervention groups, and the p values obtained from the statistical analysis have changed, the qualitative trends of statistical significance (i.e., whether students in group A performed significantly better than students in group B) remain the same as what is shown in table 4.

In order to investigate in-depth how the scaffolding support provided affected student performance, students' solutions to the friction problem were binned into categories based on their problem solving approach. Table 5 and Table 6 list students' different approaches for finding the frictional force and the corresponding percentage of students in each group in the calculus-based and algebra-based course, respectively. The p-values, which compare the difference between the number of students in the intervention groups and comparison groups who adopted different problem solving approaches, are presented in Table 7. As discussed previously, one common mistake students made was to first find the normal force by using Newton's $2^{nd}$ Law in the equilibrium situation and then using $f_s = \mu_s F_N$ to solve for the frictional force. We note that if the students' magnitude of the frictional force was correct but the overall performance on the whole quiz indicated that they were still connecting the static friction to its maximum value (for example, by using $f_s = \mu_s F_N$ to solve for the normal force in the next sub-problem after finding $f_s$ correctly), they were classified in the $2^{nd}$ category of solution approaches ($f_s = \mu_s F_N$) in Tables 5 to 7. As students' alternative approaches to the friction problem were not limited exclusively to $f_s = \mu_s F_N$, a third category named "other" was created. For example, five percent of the students confused the static friction with the kinetic friction and used $\mu_k$ instead of $\mu_s$ to solve the problem and twelve percent of the students erroneously combined $\mu_s$ and $\mu_k$ together and came up with an answer such as $f_s = \mu_s F_N + \mu_k F_N$. All these different approaches were placed in the "other" category in Tables 5 through 7.

Table 5 and Table 6 show that in all intervention groups, the percentages of students who correctly used Newton's $2^{nd}$ Law in the equilibrium situation to solve for static friction without connecting it to its maximum value were higher than those in the comparison groups in both the calculus- and algebra-based courses. With the interventions provided, the percentage of students who solved the problem correctly increased from 21% to an average of 51% in the calculus-based course, and from 15% to an average of 39% in the algebra-based course. Among all the scaffolding supports provided, interventions 2 and 3 both provided effective scaffolding in helping calculus-based students solve the static friction problem correctly, while the most effective intervention in the algebra-based course was intervention 2. The percentages of students who correctly used Newton's $2^{nd}$ Law in the direction parallel to the inclined plane to solve for frictional force in these three intervention groups were more than two times higher than that for the comparison group in the corresponding course. This result is consistent with the average scores in Table 3. We note that the increase in the percentage of students who solved the friction problem correctly would be accompanied by the decrease of the percentage of students who used either $f_s = \mu_s F_N$ or other approaches. Comparing the percentages of students in the "$f_s = \mu_s F_N$" category in particular, however, we found that only intervention group 2 in the algebra-based course showed a significant decrease from the comparison group. Although the percentages in the calculus-based intervention groups 2 and 3 also decreased by more than 10%, these differences from the comparison group were not large enough to be statistically significant for the number of students in each group. In the intervention group 1 in both courses and the intervention group 3 in the algebra-based course, the percentages of students using the "$f_s = \mu_s F_N$" approach were very similar to that in the corresponding comparison group. These findings suggest that while the three interventions provided may have helped improve student performance on the friction problem to some extent, overall, a significant component of the improvement was likely due the fact that for those students who would otherwise had great difficulty identifying information relevant for solving for static friction when no scaffolding was provided, the intervention gave them more clues about how to construct the solution to the friction problem (and therefore the percentages of students in the "other" group was reduced as shown in Table 5 and Table 6). However, in many groups, the use of $f_s = \mu_s F_N$ to solve the friction problem was still common. Overall, Table 3 to Table 7 suggest the following:

*(1) Intervention 2 was always among the most effective intervention for students in both algebra-based and calculus-based courses, while intervention 1 was always among the least effective*

*(2) Intervention 3 worked better for the students in the calculus-based course than the algebra-based course*

*(3) Even in intervention groups that are more effective (e.g., intervention 2 in both courses and intervention 3 in the calculus-based course), not all students benefited equally from the scaffolding support provided and not all of them were able to solve the friction problem correctly as intended.*

Next, we will explore students' responses to different additional tasks/scaffoldings contained in each intervention in order to gain more insight into cases in which these interventions worked or did not work, and to investigate possible ways to improve student performance on the friction problem involving an alternative conception.

**Table 5. Percentage of students in each group who used particular problem solving approaches in the calculus-based course. The average score obtained by students in each group is also listed in the parentheses for reference.**

| | Percentage of Students | | | | |
|---|---|---|---|---|---|
| | Comp | Intv 1 | Intv 2 | | Intv 3 |
| | | | Pre | post | |
| Correct use of Newton's 2$^{nd}$ Law (i.e. $f_s - mg \sin\theta = 0$) | 21.1% (9.3) | 38.2% (9.5) | 23.6% (9.2) | 56.9% (9.8) | 56.4% (10.0) |
| $f_s = \mu_s F_N$ | 42.1% (4.2) | 38.2% (4.8) | 43.1% (4.4) | 25.0% (4.8) | 30.8% (4.9) |
| Other | 36.8% (1.6) | 23.5% (1.8) | 33.3% (0.8) | 18.1% (2.3) | 12.8% (0.0) |

**Table 6. Percentage of students in each group who used particular problem solving approaches in the algebra-based course. The average score obtained by students in each group is also listed in the parentheses for reference.**

| | Percentage of Students | | | | |
|---|---|---|---|---|---|
| | comp | Intv 1 | Intv 2 | | Intv 3 |
| | | | Pre | post | |
| Correct use of Newton's 2$^{nd}$ Law (i.e. $f_s - mg \sin\theta = 0$) | 14.9% (9.0) | 30.2% (9.9) | 19.6% (9.5) | 60.8% (9.6) | 27.3% (9.8) |
| $f_s = \mu_s F_N$ | 34.0% (3.9) | 36.5% (4.4) | 41.2% (4.3) | 15.7% (4.4) | 39.4% (4.6) |
| Other | 51.1% (0.9) | 33.3% (1.4) | 39.2% (1.0) | 23.5% (1.1) | 33.3% (1.5) |

**Table 7. P-values (using the Chi-square tests) for the comparison of the number of students who used various problem solving approaches in different groups. The differences that are significant are indicated by the asterisk (\*). The pound symbol (#) indicates a marginally significant difference with a p-value between 0.05 to 0.10.**

| | | Comparison vs. intervention 1 | Comparison vs. intervention 2 | Comparison vs. intervention 3 |
|---|---|---|---|---|
| Calculus | Correct use of Newton's $2^{nd}$ Law (i.e. $f_s - mg\sin\theta = 0$) | 0.109 | 0.000* | 0.001* |
| | $f_s = \mu_s F_N$ | 0.738 | 0.065# | 0.301 |
| | Other | 0.221 | 0.029* | 0.015* |
| Algebra | Correct use of Newton's $2^{nd}$ Law (i.e. $f_s - mg\sin\theta = 0$) | 0.062# | 0.000* | 0.118 |
| | $f_s = \mu_s F_N$ | 0.789 | 0.035* | 0.562 |
| | Other | 0.061# | 0.005* | 0.059# |

2. *Intervention 1: Simply guiding students to work through the solution of the analogical problem was less effective than other approaches in helping most students*

As shown in Table 5 and Table 6, intervention 1 was less successful in helping students solve the friction problem correctly compared to intervention 2 (in both courses) and intervention 3 (in the calculus-based course). Students in intervention group 1 were asked to learn from the tension problem, and reproduce the solution to the tension problem before solving the friction problem. In order to understand if students were able to make sense of the tension problem provided and to explore where student difficulties originate from, students' ability to reproduce the solution to the tension problem was scored using the same rubric that was used for scoring the friction problem (with focus only on the part associated with "using $\sum F = 0$" approach in Table 2, since the other problem solving approaches do not apply in the tension problem.) Table 8 shows intervention 1 students' average scores on trying to reproduce the tension problem after returning its solution to the instructor before attempting to reproduce it. Students' performance on the friction problem is also listed for comparison. We found that almost every student, whether in a calculus-based or an algebra-based course, was able to solve the tension problem correctly except for some minor mistake (if any) such as confusing the weight and the mass. In particular, the average score on the tension problem was 9.7 out of 10. On the friction problem, however, the average score was significantly lower (5.9 and 5.1 in the calculus- and algebra-based courses, respectively). As Table 5 and Table 6 show, only 38% and 30% of the students, respectively, correctly used Newton's 2nd Law along the direction parallel to the inclined plane to solve for frictional force; $f_s = \mu_s F_N$ was still common in both courses after the scaffolding. We note that the rationale behind this scaffolding support is the hypothesis that students will be able to unpack the deep similarities between the analogical problems, and that if students try to solve the tension problem and the friction problem by drawing free body diagrams (FBDs) at the beginning of their solutions first, they may realize from the FBDs that the frictional force in the second problem corresponds to the tension in the first problem. However, examining students' problem solving process suggests that only less than 50% of the students drew similar FBDs for the tension problem and the friction problem. Some students didn't draw any FBD in any of the problem, some drew FBD in one problem but not in another, and some drew different FBDs for the tension problem and the friction problem (e.g., compared to the FBD for the tension problem, the FBD for the friction problem is missing a force, having an additional force such as kinetic frictional force or tension force which shouldn't be included in the friction problem, or having the frictional force in the incorrect direction). Figures 3, 4, 5 are some examples of the different FBDs students drew for the tension problem and the friction problem. If these students did not realize that the tension problem and friction problem have the same FBD, they may be less likely to benefit from the scaffolding support provided. In addition, examining the similarities written down by students suggests that the similarities most frequently noted are those that focus on the surface feature (e.g., the car has the same weight in both problems; the angle of inclination is the same; the car is at rest in both problems). Only 24% of students explicitly mentioned the similarity between the tension in one problem and the frictional force in the other problem. Some students explicitly pointed out that the tension problem is helpful for solving the friction problem because it tells them how to find the normal force, and to find the frictional force, they need to use $\mu_s F_N$. It is likely that the scaffolding support, which included asking students to identify the similarities between the two problems and reproduce the tension problem again, was not meaningful enough to engage many students in the analogical reasoning, especially if they had an alternative conception about being able to solve the friction problem using a different approach. Therefore, their improvement with intervention 1 was not as good as some other intervention(s) in which students were provided with more direct scaffolding and hints to help them contemplate the applicability of the equation $f_s = \mu_s F_N$ carefully in the friction problem.

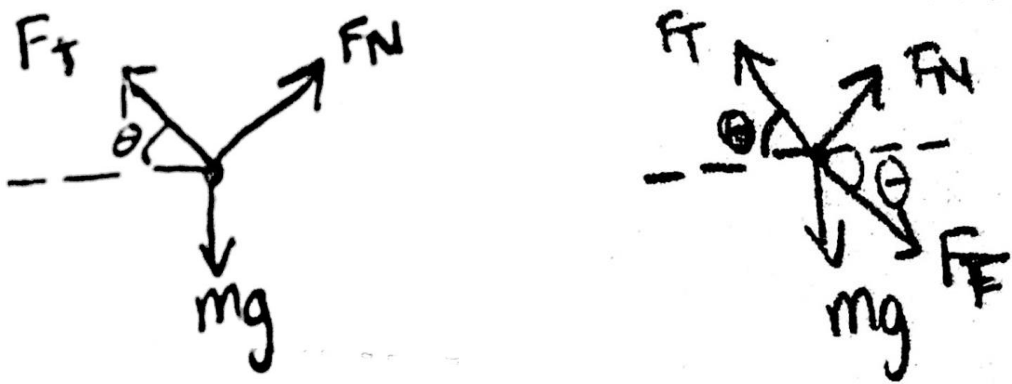

**Figure 3. An example of the different FBDs drawn for the tension problem (left) and the friction problem (right) by the same student. This student made two mistakes in the FBD for the frictional force: (1) An additional tension force was mistakenly included (2) The direction of the frictional force was incorrect.**

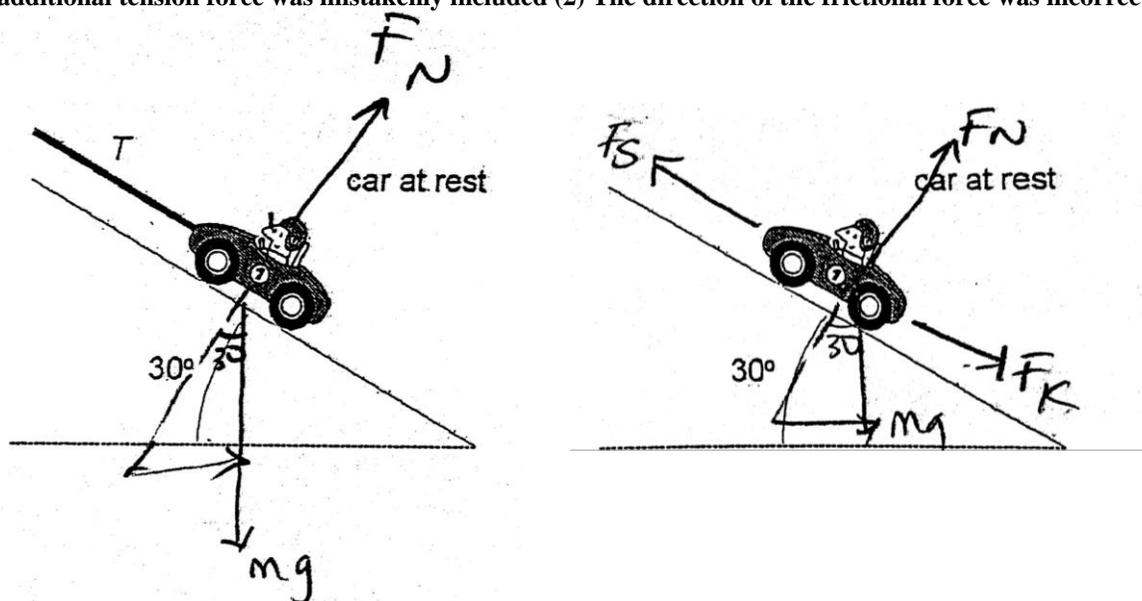

**Figure 4. An example of the different FBDs drawn for the tension problem (left) and the friction problem (right) by the same student. This student mistakenly included an additional force – the kinetic friction – in the FBD for the friction problem.**

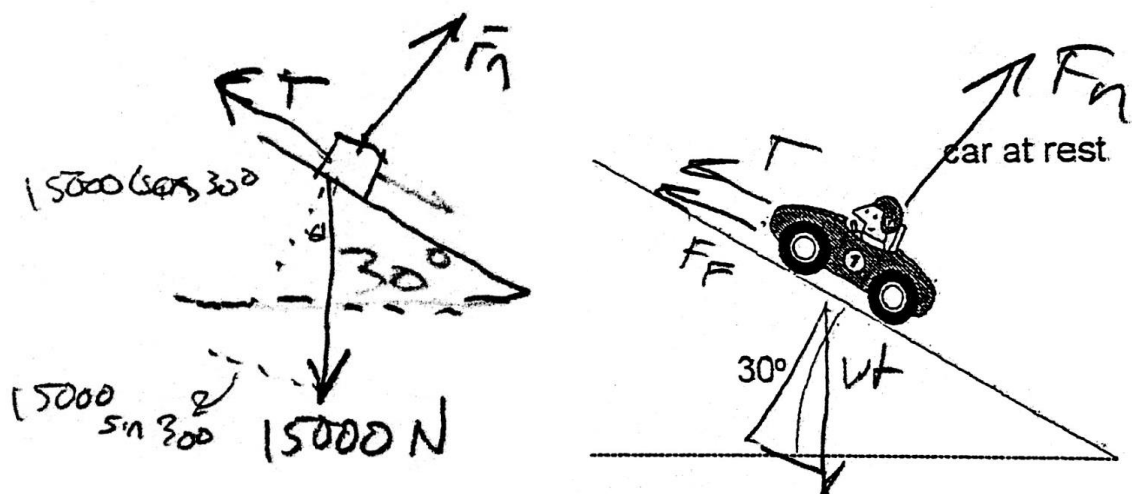

**Figure 5. An example of the different FBDs drawn for the tension problem (left) and the friction problem (right) by the same student. This student mistakenly included an additional tension force in the FBD for the friction problem.**

**Table 8.  Average scores out of 10 points on the tension problem (reproduced after returning its solution) and the friction problem for intervention 1 in the calculus-based and algebra-based courses.**

| Tension Problem | | Friction Problem | |
|---|---|---|---|
| Calculus (N=34) | Algebra (N=63) | Calculus (N=34) | Algebra (N=63) |
| 9.7 | 9.7 | 5.9 | 5.1 |

3.      *Intervention 2 was generally the most effective for students in both calculus-based and algebra-based courses*

As noted earlier, intervention 2 was always one of the most effective interventions for students in both the calculus- and algebra-based courses. Students in this group were asked to solve the friction problem on their own before learning from the analogical solved example involving tension. Moreover, they were advised to make a qualitative prediction about the magnitude of the static frictional force on a steeper incline based on their daily experience and compare their prediction with their calculated result. It is possible that the scaffolding support in this intervention is more helpful for students because of the clear targeted goal and the thinking process students went through in their first attempt to solve the friction problem, which may facilitate better transfer from the solved tension problem later. Similar findings showing an advantage in postponing scaffolding until students have attempted to solve the problem without help have been discussed in other situations [53,54]. In addition, the fact that intervention 2 involves additional questions about the steeper incline that guided students to devote effort to explicitly contemplate their solutions to static friction could have also contributed to the improvement in student performance. As Table 5 and Table 6 show, in both the algebra-based and calculus-based courses, intervention 2 helped decrease the percentage of students who used the $f_s = \mu_s F_N$ approach by 17~18%. While encouraging, Table 5 and Table 6 also indicate that there are still 16%~25% of students who used the $f_s = \mu_s F_N$ approach in intervention group 2. In order to understand why not all students benefited from the scaffolding support provided as we intended, students' responses to those additional questions about the steeper incline were examined.

We found that in both courses, most students' reasoning behind their first predictions about the magnitude of the static frictional force on a steeper incline could be classified into one of three categories: (1) daily experience and correct interpretation/prediction, (2) daily experience and incorrect interpretation/prediction, and (3) answer based on the calculated result. There were students from both the calculus-based and algebra-based courses who were able to connect the problem with their daily experience and make a correct prediction. For example, a calculus-based student correctly stated that "*Based on my daily experience, I would predict that the magnitude should be larger because the steeper angle makes objects want to move more than the slight angle*". Similar statements such as "*If the inclined plane is steeper, the frictional force between the object and the surface will be larger because the frictional force is equal to the magnitude of the force pulling you down the incline (just in the opposite direction) and from daily experience it feels like more force is trying to pull you down a plane when the plane is steeper*" were made by some algebra-based students as well. However, although students knew from their daily experiences that it is less likely for an object to stay at rest on a steeper incline, some of them had an alternative explanation so that their prediction was opposite to their intuition. For example, one student said "*Based on my daily experience, the frictional force should be less on a larger incline because it's harder to stay at rest on a steeper incline.*" Such explanations were found in both the algebra- and calculus-based courses. We note that the purpose of this prediction question was to help students who originally adopted the $f_s = \mu_s F_N$ approach to discover the conflict between the qualitative trend suggested by the daily experience (static friction should be larger on a steeper incline) and their quantitative answer (showing that the static friction calculated using $f_s = \mu_s F_N$ is smaller on a steeper incline). We hypothesized that such questioning will provide incentive for students to re-examine their problem solving approach if they used $f_s = \mu_s F_N$ to solve for friction in their first attempt. However, as indicated by the data, not all students were able to discover the inconsistency in their responses. Some students provided alternative explanations about their daily experience as in the example described above, some made a prediction not based on their daily experience but based on a quantitative calculation, and some predicted the correct outcome using the correct reasoning but made a mistake in the subsequent calculation so that the frictional force they calculated using $f_s = \mu_s F_N$ happened to be larger on a steeper incline, which is consistent with their qualitative prediction. In total, we found that only 14% of the students who used $f_s = \mu_s F_N$ in their first attempt to solve for static friction discovered the inconsistency in the manner it was originally intended. It is possible that intervention 2 could have a larger impact on student performance than it currently does if the scaffolding support provided can be modified so that more students are able to notice the deficiency in using $f_s = \mu_s F_N$ and re-examine their problem solving approach.

4.    *Intervention 3 was more helpful in the calculus-based course than in the algebra-based course and many students did not discern the full meaning of the inequality*

As for intervention 3, which not only provided students with the solved tension problem but also exposed them to the correct inequality $f_s \leq \mu_s F_N$ and asked them to explicitly discuss whether $\mu_s$ is needed to solve the friction problem by thinking about the meaning of the inequality, Table 6 suggests that this scaffolding support was more beneficial to the students in the calculus-based course than the algebra-based course. In order to get more insight into how students responded to the scaffolding support in this intervention and why there was a big difference between the calculus-based and the algebra-based courses, the percentages of students who explicitly answered whether $\mu_s$ is needed/not needed in these two courses were examined. Table 9 suggests that even though students were advised to identify the similarity between the two problems and were also explicitly shown that the correct expression for the static friction was not $f_s = \mu_s F_N$ but $f_s \leq \mu_s F_N$, some students (especially many of those in the algebra-based course) had difficulty in making sense of the inequality and its implication for the friction problem. As Table 9 shows, fifty percent of the students in the algebra-based course explicitly noted that in order to find the frictional force on the car, $\mu_s$ needs to be given. Examining students' explanations of the inequality, we found that many students were not able to take advantage of the scaffolding provided because they focused only on one aspect of the inequality and failed to see its full implication: Instead of realizing that "$f_s$ can be any value from zero to the maximum value (which is $\mu_s F_N$), depending on how strong the opposing force is", they only focused on the fact that static friction cannot be larger than $\mu_s F_N$. They explained that if this maximum amount is exceeded, the object could no longer be stationary; however, since the car in the problem was at rest, the coefficient of static friction must be used. The similarities between the two problems and explicitly asking students to explain the inequality sign did not help them realize that the static frictional force in the friction problem was not equal to its maximum value. In addition, there were some students who literally wrote out $f_s \leq \mu_s F_N$ in words as their explanation of the inequality. If these students did not discern deeply what the inequality means, they were less likely to benefit from the scaffolding support provided. We also found that a few students incorrectly interpreted the inequality and the maximum static friction. For example, one student stated that "$f_s \leq \mu_s F_N$ *means that the normal force multiplied by* $\mu_s$ *must be greater than fₛ in order for the car to overcome the frictional force. If it is not greater, then the car will not move.*" Another student noted "*It is an inequality because if* $\mu_s F_N$ *were any smaller than fₛ, it would cause the force to be too small and the car would move.*" Such difficulty in interpreting the relationship between the static frictional force and its maximum value was more commonly found in the algebra-based course than in the calculus-based course. It is likely that the scaffolding support provided in intervention 3 requires an ability to interpret inequalities at a level which is more suitable for students in the calculus-based course but too innovative for many students in algebra-based courses. Therefore, intervention 3 may be more commensurate with students' prior skills in a calculus-based course but may be beyond the zone of proximal development [62] for many students in an algebra-based course. Accordingly, more students from calculus-based course benefited from this scaffolding support than those in the algebra-based course.

**Table 9. Percentage of students in intervention group 3 who answered that $\mu_s$ is needed/not needed in the friction problem after they attempted to explain the meaning of the inequality $f_s \leq \mu_s F_N$.**

|  | Calculus (N=39) | Algebra (N=66) |
|---|---|---|
| $\mu_s$ not needed | 69.2 % | 45.5 % |
| $\mu_s$ needed | 28.2 % | 50.0 % |
| Irrelevant answer or no answer | 2.6 % | 4.5 % |

5.      *The analogical problem solving activity and related scaffolding supports provided played an important role in the effectiveness of an intervention. Simply increasing the time and information provided to students did not improve their performance.*

Before ending this section about the effects of different scaffolding support in calculus-based and algebra-based introductory physics courses, we note that the better performance observed in intervention group 2 (in both courses) and intervention group 3 (in the calculus-based course) came from the analogical problem solving and the related scaffolding supports provided, not simply from the longer time and more information given to students in these intervention groups compared to the comparison group. While students in these intervention groups were given more total time than those in the comparison group (students in the comparison group did not spend time browsing over the solution to tension problem or contemplating the answers to the additional questions asked to scaffold their learning), simply giving students more time and more information would not necessarily result in a better performance as great as what we observed in this study. In order to confirm this hypothesis, we examined the performance of another group of students in an equivalent algebra-based course at the same institution who were comparable to the algebra-based students in the study discussed based on their FCI scores. This other group of students in the algebra-based course (N=22) was given 30 minutes to work on the friction problem in a mandatory quiz. Similar to the comparison group, no solved tension problem was provided to students in this group. However, in order to encourage students to think about the problem for the full 30 minute period, they were instructed to devote at least the first 10 minutes to think about how to solve the friction problem before actually writing down their solutions. Moreover, they were allowed to use their textbooks and notes during the full 30- minute quiz period. Students' performance on this quiz was graded using the same rubric presented in Table

2. We found that although students in this group had 30 minutes to work on the quiz (the maximum amount of time students in the intervention groups received)) and they could use all the information that they found from the textbooks and notes to solve the problem, their performance (with an average score of 4.6 out of 10) was still significantly lower than the performance of intervention group 2. In fact, their performance was not statistically different from the performance of the comparison group discussed earlier. Similarly, when a group of students in an equivalent calculus-based course (N=38) was given 30 minutes to work on the friction problem with the help of their textbooks and notes (but none of the analogical problem solving and scaffolding used in the intervention groups were provided), they achieved a score of 3.7 out of 10 on the friction problem, which is statistically lower than the score achieved in the calculus-based intervention groups 2 and 3 (even though their FCI scores were not statistically significant). Such findings suggest that simply providing more time and more information to the students does not guarantee a better performance unless the scaffolding support provided engages students in a deep cognitive processing about their alternative conception and approach to problem solving.

## B.      Findings from the individual interviews

### 1.      General Description of the interviews

The quantitative data suggest that while some of the scaffolding supports provided in the interventions helped improve student performance on the friction problem, overall, there is still room for improvement. In particular, even though students in intervention groups 2 and 3 received scaffolding support and extra hints to help them reconsider their alternative conceptions, employing $f_s = \mu_s F_N$ to solve the problem was still common. In order to understand the cognitive mechanism for how students were impacted by the scaffolding supports and to further explore strategies to help students, ten student volunteers from other equivalent introductory physics classes who did not participate in the quiz were recruited for one-on-one interviews the following semesters. From these semi-structured think-aloud interviews, we can obtain a better understanding of students' thought processes while they solve the problem with different scaffolding supports. At the time of the interview, three interviewees were concurrently enrolled in the $1^{st}$ semester introductory physics course, and seven interviewees had finished their $1^{st}$ semester introductory physics course in a previous semester. In total, five out of the ten interviewees took the calculus-based mechanics course, and the other five of them took the algebra-based mechanics course. The interviews were conducted after all the relevant topics had been covered in the lectures, and there was no qualitative difference found between students who were concurrently enrolled in the introductory mechanics course and those who had finished the course.

During the interviews, students were asked to learn from the solved tension problem provided and solve the analogical friction problem. Three students in the algebra-based course and two in the calculus-based course were provided with the scaffolding support of intervention 2. The other interviewed students were given the scaffolding support of intervention 3. Students were asked to perform the task while thinking aloud; they were not disturbed during the task. After the students completed the task, the researcher first asked clarification questions in order to understand what they did not make explicit earlier and what their difficulties were. Based on this understanding, the researcher then provided additional support to the students in order to help them solve the friction problem correctly if they had not done so. At the end of the interviews, students were invited to reflect on the learning process they just went through and provide some suggestion from their own perspective on how to improve students' performance on the friction problem. The goal of the students' reflection was to help us identify possible helpful scaffolding support not only based upon what the researchers observed but also based upon students' reflections of their own learning. The interview protocol can be found in the auxiliary materials. All the interviews were recorded, and additional observations from the interviews were immediately noted down right after each interview was completed. Results from the interviews were analyzed based on students' written solutions, the researcher's notes and the recording. In particular, we focused on identifying what difficulties students encountered when attempting the task, and what scaffolding support appeared to benefit the students. Two researchers discussed how to interpret the observations from the interviews together until they reached an agreement. The findings are presented below.

2.      *Interviewed students struggled with the idea that static friction does not have to equal its maximum value*

There are two major findings from the interviews. First, the interviews revealed that many students struggled with the idea that the static friction is not necessarily equal to its maximum value. Even if students were able to learn from the scaffolding provided and discern the similarities between the tension force in one problem and the frictional force in the other problem, often they were still confused because they did not know why $f_s$ does not have to equal $\mu_s F_N$. For example, when a student (student A) who was provided with intervention 3 in the interview attempted to use both Newton's $2^{nd}$ Law in the equilibrium situation along the direction parallel to the inclined plane and the equation $f_s = \mu_s F_N$ to solve for friction, he was confused when he saw that different methods yielded different answers. He was not sure if he should say yes or no to the question related to "whether $\mu_s$ is needed in order to solve for friction." For example, the student continued "*Here, the question is: can you find the frictional force on the car in the problem without knowing the coefficient of static friction? I would say no, but my equation says yes. That doesn't make sense. Because, judging by my free body diagram[shown in figure 6], $mg\sin\theta$ would actually…would have to equal force done by friction. But I thought the definition of the force done by static friction was the coefficient of friction times the normal force.*" This student also explicitly noted that he did not know how to explain the meaning of the inequality. He knew that the inequality stated that the static friction is smaller than or equal to the coefficient of static friction times the normal force, and that the static friction could not be larger, but he did not know how to interpret the inequality. In particular, he noted that he did not know why $f_s$ will ever be smaller, and not just equal to $\mu_s F_N$.

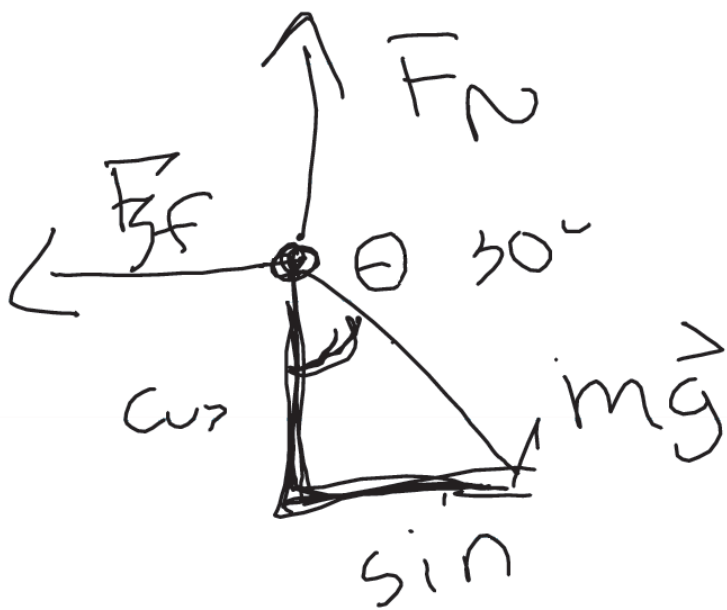

**Figure 6. Student A's FBD for the friction problem**

The same difficulty was commonly found in other interviewees, too. For example, another student, student B, who was given intervention 2, wrote down correct answers to all of the questions without using $f_s = \mu_s F_N$, although she claimed during the "thinking aloud" process that the static friction and the normal force she calculated would be connected by the formula $f_s = \mu_s F_N$. (Her free-body diagram is shown in figure 7). When the researcher later asked her to check this relation by substituting the numbers she obtained, she found a conflict and did not know what to do about it. Although learning from the solved problem made her confident that her original answer $f_s = mg \sin \theta$ was correct, she did not understand why the static friction does not have to equal $\mu_s F_N$. These findings suggest that in addition to those reasons identified in the quantitative study, another reason why the interventions are not more effective may be that they focused mainly on helping students recognize that the static frictional force does not equal $\mu_s F_N$, but did not provide enough support in helping students understand why the static frictional force can be smaller than $\mu_s F_N$. For the interventions to be more effective, more scaffolding support that focused on the latter issue may be required.

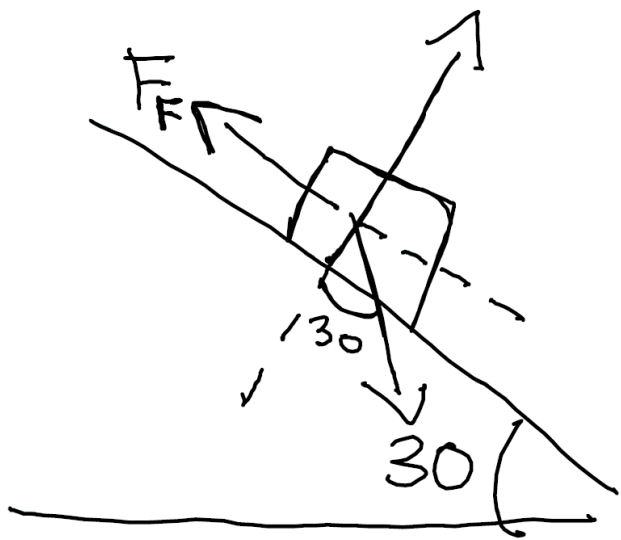

**Figure 7. Student B's FBD for the friction problem**

The second finding from the interviews (such as the one with student B) is that even if a student's written solution for the static frictional force and normal force does not involve $f_s = \mu_s F_N$, it does not necessarily indicate that they thought that this relationship does not hold in the given situation. These findings from the interviews suggest that the quantitative data presented previously could be considered as an upper limit for how well the interventions helped students overcome their reasoning difficulty by asking them to discern the similarities between the tension problem and the friction problem. Some students may require more scaffolding support in order to re-organize their knowledge about static friction and interpret the associated inequality correctly. From the interviews, we identified some discussion threads that were useful in helping students build a better understanding of static friction. We will discuss some of them in the following subsection.

### 3.    *Additional scaffolding supports that may help students*

First of all, the interviews suggest that in order to improve students' understanding of the inequality $f_s \leq \mu_s F_N$, one helpful strategy is to quantitatively lead students to reason about the static friction acting on an object (such as a heavy desk) placed on a horizontal surface and how this force keeps increasing when we push the object harder and harder until the maximum static friction is reached and the static frictional force is no longer able to hold the object in place, at which point the object starts to move. We note that although the information contained here is similar to that in intervention 2 when the angle of inclination is increased, the example of an object on a horizontal surface may be easier for students to comprehend because it does not require a decomposition of forces and the normal force stays the same. We also note that although a similar example is often used by many instructors while lecturing, it is likely that students pay more attention to one aspect of the inequality (i.e., the static friction cannot be larger than $\mu_s F_N$ otherwise the object will start to move) than the other (the situations in which $f_s$ is not equal to but smaller than $\mu_s F_N$) as we discussed in the quantitative results section. It would therefore be beneficial if additional scaffolding supports were provided to guide students to reason about this latter aspect by starting with some cueing questions. For example, students could be asked: "Since the inequality $f_s \leq \mu_s F_N$ implies that the static friction can be smaller than (not equal to) $\mu_s F_N$, can you think of any situation in which $f_s$ is indeed smaller?" or "what would happen if the static frictional force is always equal to $\mu_s F_N$?" If the students struggle with the former question, we could guide them to think about the static friction acting on an object resting on a horizontal surface when no horizontal force is applied, which many students are able to answer immediately, so that students can build their understanding on a solid base in a familiar situation. Interviews suggest that although simply asking students to consider the case in which there is no static friction while the normal force is nonzero may not be enough to totally clear up student' confusion about whether it is correct to solve for the friction without using $\mu_s$, it provides a good starting point. For example, after successive follow up questions in which student A was asked to calculate the static frictional force on a desk on the floor when 1 N, 2 N…. of external horizontal force is applied until the value of $\mu_s F_N$ was exceeded and the object starts to move, he gradually understood the full implication of the inequality and he was no longer perplexed by the fact that his answer to the friction problem did not involve the coefficient of static friction. When the researcher later asked him to reflect on his learning during the whole activity and identify the support which he found most helpful, he said "*I think it was the…the analogy of the desk [that really helped me]. I understand that now. Because…like….at first it [$f_s \leq \mu_s F_N$] just looks like an equation to me. But after I understand that it's gonna be*

*LESS THAN or equal to until that point where you exceed it, and it starts moving the other direction that's gonna be greater than, that makes sense to me.*"

Similarly, when another student was asked at the end of the interview to provide some suggestions on how to help students learn that $f_s$ is not always equal to $\mu_s F_N$ by reflecting on her learning process during the activity, she pointed out that the example of an object on a horizontal surface was very helpful, especially the part in which the researcher guided her to examine what will happen if static friction was always equal to $\mu_s F_N$. By drawing the free body diagram, this student was able to reason that if the static frictional force had a fixed value of $\mu_s F_N$, giving the object a small push toward the left (with a magnitude smaller than $\mu_s F_N$) would result in a freak phenomenon of the object moving toward the right (since a static friction acting toward the right should exist to resist the tendency to move and its larger magnitude suggests that the object would move in a direction opposite to the direction in which it was pushed), which was contradictory to her daily experience. She claimed that this example helped her the most in realizing that $f_s = \mu_s F_N$ is only true in special situations and she was able to reason about it using the free body diagram. Similar discussions about the contradictory consequences of assuming that the static frictional force is always equal to $\mu_s F_N$ were found to be very helpful for other students as well. For example, one student (student C) who reasoned about a similar contradictory consequence in the context of the friction problem had the following conversation with the researcher:

Student C: "*I never thought of it that way and that [if $f_s = \mu_s F_N = 11691\,N$] would actually be pulling it[the car] up. I am thinking it's just the…it's like money in the bank. I got more money in the bank up here and less money in the bank up there. But I have more money in the bank up here. It keeps the car from going downhill. But in effect, right, what you are saying is true. It would then have to in effect pull the car uphill.*"

Researcher: "*Right, so you don't have to pay so much money to keep the car at rest.*"

Student C: "*Right*"

Researcher: "*You just pay whatever you need*"

Student C: "*Pay the amount, right, that's equal to what force is pulling it down. Right… and I'm seeing it as a…as not a force that's pulling it, but just a total amount, like, you know, his muscle was strong enough to lift 300 pounds, but he was lifting 150 pounds, that kind of* *thing, where in essence the static frictional force is gonna not be more than its opposing …than its opposing force. So it never kind of hit me until now*"

Regarding students' suggestions for strategies that they personally found helpful for learning about static friction, many of them discussed the importance of emphasizing the inequality of static frictional force. For example, one student pointed out that "*it probably would have been better if you would say $f_s$ , the frictional force of static friction, does NOT equal $\mu_s F_N$ like 99 % , like 90% of the time or 99% of the time. But in those few instances where… and whatever those instances would be… 10 % or 5% of cases of problems…it is gonna equal it." "But then when it equals it, that's the max."* Although the concept of "static friction not always equal to $\mu_s F_N$" is usually discussed in a typical introductory physics course with the help of an equation $f_s{}^{max} = \mu_s F_N$ or the inequality $f_s \leq \mu_s F_N$, our study indicates that a majority of students were not able to grasp the full meaning of these expressions. For some students, the expression in terms of $f_s{}^{max} = \mu_s F_N$ may be more helpful than $f_s \leq \mu_s F_N$ because the former emphasizes the fact that it is the MAXIMUM static friction that equals $\mu_s F_N$ whereas the latter expression is confusing and may come across as an equality (rather than an inequality). Other students, on the other hand, may not fully understand why the former expression involves "max" and simply treat it as $f_s \; = \mu_s F_N$. One interviewee suggested that writing both $f_s{}^{max} = \mu_s F_N$ and $f_s \leq \mu_s F_N$ together would help students understand the concept of static friction better . She pointed out that "*It's easy if you put that 'f_s max would equal' equation and the inequality, and then you were to explain why. That would help a lot as to why $\mu_s F_N$ doesn't equal [the static friction]*." She reflected on her own learning of the subject and said: "*Because when my professor first taught it to us, he wrote maximum and I was like, 'hey…what does that mean?' Like… you know…it was just max, whatever. Now that we went over this, I do understand why he would put that, and I've grasped the concept better.*" Based on this student's suggestion, it is likely that listing both the equation and the inequality together would help students focus on both aspects of the inequality when the maximum static friction is is not reached. This student's remarks also emphasized that if the analogical reasoning activity as well as other scaffolding support provided were designed and implemented in a way that is commensurate with students' current knowledge, they are more likely to benefit from it.

# IV.     SUMMARY

In this investigation, we examine the effects of using analogical problem solving and different scaffolding supports to help 410 introductory physics students in algebra-based and calculus-based courses solve a problem that typically involves a strong alternative conception. The findings suggest that some students were able to take advantage of the scaffolding supports provided and transfer their learning from the solution to the tension problem provided to solve the analogical problem involving friction. In particular, after learning from the scaffolding provided, more students were able to identify the relevant concepts involved in solving the friction problem, and the score on average improved in all but one intervention group. However, the alternative conception that "static friction is always equal to its maximum value $\mu_s F_N$" was still prevalent.

Out of the three interventions used, intervention 1 did not explicitly address the alternative conception that $f_s$ is always equal to its maximum value $\mu_s F_N$. The rationale for intervention 1 was that students may benefit from the similarities in the two problems and employ Newton's 2nd law in the direction parallel to the inclined plane to solve the friction problem. No other scaffolding support to guide them to contemplate over their alternative conception related to static friction was provided. The fact that in both algebra-based and calculus-based courses, about the same percentage of students in intervention group 1 used the "$f_s = \mu_s F_N$" approach as in the comparison group to solve for the friction force suggests that for this type of problem involving a strong alternative conception, scaffolding support that addresses such difficulty more directly (such as intervention 2 or 3) is necessary to help students.

Our results also suggest that the effects of the scaffolding supports provided depend not only on the design of the scaffolding itself, but also on how closely the scaffolding matches students' prior knowledge and skills. For example, intervention 3 was found to be more helpful in the calculus-based course than in the algebra-based course. Although the scaffolding support in intervention 3 may be considered as a big hint by instructors since students were explicitly guided to the correct expression for static friction, not all students were able to benefit from it. Many students focused on the fact that the static friction could not be larger than $\mu_s F_N$ while trying to explain the inequality, but ignored the fact that the inequality suggests that static friction can in fact be *smaller* than $\mu_s F_N$. Interviews suggest that in order to make sense of the scaffolding support in intervention 3, students need to correctly understand and be able to explain the mathematical inequality provided. Many interviewed students struggled to explain what the inequality means and could only articulate that the static frictional force cannot be larger than the maximum value. Many of them explicitly expressed confusion when the interviewer asked whether the static friction can be smaller than its maximum value in a given situation. The fact that not as many students in the algebra-based course benefited from intervention 3 as in the calculus-based course suggests that this scaffolding support may be more commensurate with calculus-based students' knowledge and skills.

Overall, we found that intervention 2, in which students had to compare their predictions from daily experiences with their calculations and solve the friction problem first before the tension problem was provided was consistently among the most effective in helping students refrain from using $f_s = \mu_s F_N$ and instead use Newton's 2nd law along the direction parallel to the inclined plane to find static friction in both the algebra- and calculus-based courses. This result suggests that if an instructor wants to use a similar analogical problem solving activity with additional scaffolding support to help students solve a target problem that typically involves a strong alternative conception, it is a good strategy to provide the solved analogical problem to students only after they have tried to solve the target problem (e.g. the friction problem) on their own first. Providing students with additional scaffolding, e.g., in the form of questioning that help students recognize and resolve the conflict between their alternative conception and their other knowledge elements (e.g., knowledge drawn from everyday experiences) can also be helpful.

While some level of success was found in intervention 2 and 3 which provided more explicit scaffolding than intervention 1, our study has also identified possible strategies for future improvement in problems involving strong alternative conceptions. First, it may be helpful if the scaffolding support provided in these problems can also guide students to make sense of the scientific concepts they are struggling with in more depth. The scaffolding support used in our study helped some students, but did not place enough emphasis on helping others understand why the approaches suggested by the analogies are more appropriate for solving the problem. Interviews suggest that even though some students understood that the scaffolding guided them to use Newton's $2^{nd}$ law along the direction parallel to the inclined plane to solve for friction instead of using "$f_s = \mu_s F_N$", they were still confused because they did not know why $f_s$ will ever be smaller than $\mu_s F_N$ or why $f_s$ does not have to equal $\mu_s F_N$. If additional scaffolding is included to help students consider cases in which the same content is involved but invoking alternative conception more directly causes a cognitive conflict (e.g., there is zero frictional force acting on a book resting on a desk, even if $\mu_s$ and $F_N$ are both nonzero), more students are likely to benefit. Second, it may be helpful to increase the number of examples provided in which students are provided scaffolding support to realize the conflict associated with alternative conceptions. If the scaffolding support involves only one such example, students may not necessarily discern the need to reconsider their conception deliberately and repair, reorganize and extend their knowledge structure. With more guided examples with gradually reduced scaffolding support, students are more likely to develop self-reliance and restructure their knowledge. In addition to the scaffolding supports used in the classroom investigations discussed, other scaffolding that can be provided include guiding students to think about the frictional force acting on a stationary book on the horizontal desk, guiding them to consider what would happen if the static friction always equals its maximum value, as discussed with the interviewed students in this study. In addition, it may also be helpful to ask students to consider a new case, which combines both the tension problem and the friction problem with a car held at rest with a rope on a frictional incline. In this case involving both tension and friction, depending on how strong the tension force is, the magnitude of the static frictional force can be any value between zero Newton to $\mu_s F_N$, which might provide another strong case to demonstrate the inequality of static friction. Future studies could explore the effects of these interventions (and possibly a combination of several different interventions) to identify more effective strategies for helping students. Future studies can also explore how these types of interventions with different scaffolding supports help students solve other quantitative problems involving alternative conceptions.

Finally, we re-emphasize the importance of systematically studying the effects of scaffolding supports on problem solving involving problems in which alternative conceptions can derail the entire process. The design of an effective instructional intervention is a challenging task and a scaffolding support which appears promising based upon a cognitive task analysis may turn out not to be as effective as anticipated for a particular student population. It will be valuable for instructors looking for strategies to assist students with such problems if the effectiveness of a variety of scaffolding supports is examined across problems.

# V.  ACKNOWLEDGEMENTS

We thank the National Science Foundation for support. We are very grateful to professors F. Reif and R. P. Devaty for extremely helpful discussions and/or feedback on the manuscript.